\documentclass[english,pra,reprint, aps, showkeys, showpacs, superscriptaddress]{revtex4-1}
\usepackage[LGR,T1]{fontenc}
\usepackage[latin9]{inputenc}
\usepackage{float}
\usepackage{amsmath}
\usepackage{amssymb}
\usepackage{stackrel}
\usepackage{graphicx}
\usepackage{mathrsfs}
\usepackage{color,xcolor}
\makeatletter

\pdfpageheight\paperheight
\pdfpagewidth\paperwidth

\ProvideTextCommand{\~}{LGR}[1]{\char126#1}

\usepackage[colorlinks,citecolor=blue,urlcolor=blue]{hyperref}
\usepackage[colorlinks]{hyperref}
\hypersetup{%
	plainpages=true,
	breaklinks=true,
	hypertexnames=false,
	pageanchor=true,
	colorlinks=true,
	linkcolor={blue},
	citecolor={red},
	urlcolor={blue},
	anchorcolor={black}
}

\renewcommand{\eqref}[1]{\mbox{Eq.~(\ref{#1})}}

\begin{document}
\title{Topological State Transfer through Effective Boundary-Mode Channels}
\author{Chong Wang}
\affiliation{Guangdong Provincial Key Laboratory of Quantum Metrology and Sensing
	\& School of Physics and Astronomy, Sun Yat-Sen University (Zhuhai
	Campus), Zhuhai, China}
\affiliation{School of integrated circuits, Tsinghua University, Beijing, 100084,
	China}
\author{Xiu Gu}
\affiliation{School of integrated circuits, Tsinghua University, Beijing, 100084,
	China}
\affiliation{Frontier Science Center for Quantum Information, Beijing, China}
\author{Shu Chen}
\affiliation{Beijing National Laboratory for Condensed Matter Physics, Institute of Physics, Chinese Academy of Sciences, Beijing 100190, China}
\author{Yu-xi Liu}
\email{yuxiliu@mail.tsinghua.edu.cn}
\affiliation{School of integrated circuits, Tsinghua University, Beijing, 100084,
	China}
\affiliation{Frontier Science Center for Quantum Information, Beijing, China}
\date{\today}

\begin{abstract}

Localized edge modes provide a compact channel for transferring an excitation through an extended lattice, but the relation between the microscopic chain and the few levels that actually govern the transfer is often left implicit. We develop this boundary-subspace description for a superconducting-qubit realization of the Rice--Mele model in the single-excitation sector. For one finite chain, projection onto the two edge modes yields a Landau--Zener Hamiltonian whose detuning and hybridization identify the adiabatic transfer path and the spectral constraint relevant to protocol optimization. We then join two Rice--Mele segments at a single physical boundary qubit. The resulting three localized modes form an effective three-state channel with microscopically determined, independently tunable energies and couplings. This channel supports both a zero-energy dark-state passage and an open Rice--Mele-type passage driven by time-dependent bonds and on-site potentials. Direct simulations of the complete qubit chains reproduce the effective-channel dynamics. Under matched control bounds, dividing the transport distance at a shared boundary also preserves larger finite-size edge-mode couplings and reaches a stable high-fidelity regime faster than transfer through one uninterrupted chain. The boundary-mode projection thus relates the microscopic controls directly to the effective energies, finite-size couplings, and adiabatic gaps that govern the transfer.

\end{abstract}

\maketitle

\section{Introduction}

Transferring a quantum excitation across a controllable many-body device is both an information-processing primitive and a test of coherent lattice dynamics. One strategy is to engineer the couplings of an entire chain so that its spectrum performs the desired mapping~\cite{Christandl2004Perfect,Bose2003Quantum,Petrosyan2010State,Gualdi2008Perfect}. A different strategy is to encode the excitation in localized boundary modes and steer only the small low-energy sector formed by those modes. The latter viewpoint is especially useful in finite topological chains: it connects the spatial structure of an edge state to an effective few-level avoided crossing and makes the energy scales that limit adiabatic transfer explicit.

The Su--Schrieffer--Heeger (SSH) chain is the canonical one-dimensional setting for such boundary modes~\cite{Su1979,Su1980,Heeger1988,Asboth2016}. Its alternating bonds distinguish two dimerization patterns, one of which supports end-localized states in an open geometry. The same physics has been investigated in photonic and acoustic structures, cold atoms, and other engineered lattices~\cite{Lu2014a,Chien2015,Goldman2016,zhu2018Topological,Poli2014,Kitagawa2011,Cardano2017,Atala2013,Meier2016}. Adding a staggered on-site potential gives the Rice--Mele model and supplies a second control coordinate with which an instantaneous boundary eigenstate can be continued from one end of a finite chain to the other. Periodic modulation of this model also underlies the Thouless pump~\cite{Thouless1983,Lohse2015,Nakajima2016}. The process considered here, however, is an open adiabatic path between edge states; it need not form a closed parameter cycle and is therefore distinct from Chern-number-quantized pumping.

For a long finite chain, the states involved in this open transfer occupy only a small portion of the full Hilbert space whenever they remain spectrally separated from the bulk. Projecting onto the left and right edge modes then produces a two-state Hamiltonian: the boundary-energy difference acts as a detuning, while finite-size overlap supplies the off-diagonal coupling. This is precisely the structure of a Landau--Zener (LZ) problem~\cite{landau2013quantum,yatsenko2002topology}. It provides an economical explanation of the transfer and also exposes its central limitation. Stronger edge localization improves separation from bulk states but suppresses the end-to-end matrix element exponentially, so localization, spectral isolation, and transfer time cannot be optimized independently.

Superconducting quantum circuits are well suited to examine this boundary-channel picture. Their transition frequencies and exchange couplings can be designed and tuned, and arrays of coupled qubits can represent lattice sites directly~\cite{You2011,Devoret2013,Gu2017,Gu2017a,yuyang2019Experimental,xue2016topological}. Geometric and topological properties have already been accessed with superconducting circuits~\cite{Leek2007,Berger2012,Berger2013,Schroer2014,Zhang2017,Roushan2014,Flurin2016,Ramasesh2017}. Here we use a chain of gap-tunable flux qubits~\cite{Paauw2009,Paauw2009Thesis,Schwarz2013,zhu2010coherent}; conservation of excitation number maps its single-excitation sector onto an SSH or Rice--Mele tight-binding chain, with qubit frequencies providing on-site potentials and qubit--qubit exchange providing the bonds.

Our first objective is to derive the effective boundary dynamics of a single Rice--Mele chain and use its two-state LZ structure to interpret and improve end-to-end transfer. Our second objective is to turn this reduction into a design principle. We identify the right edge of one Rice--Mele segment with the left edge of a second segment, so that the junction is one shared physical qubit rather than two sites joined by an extra link. Projection now yields a three-state Hamiltonian for the left, shared, and right boundary modes. We derive its matrix elements from the microscopic chains and use the resulting channel in two ways: a counterintuitive coupling sequence follows a zero-energy dark state, whereas coordinated bond and on-site-potential modulation follows a nonzero-energy open Rice--Mele-type branch. Full-chain calculations test the range of the three-mode description. A resource-matched length comparison further shows why the shared node can accelerate transfer: it replaces one exponentially small overlap across the entire distance by larger overlaps across two shorter segments.

\section{Topological qubit chain with superconducting qubits}\label{sec:qubitchain}

\begin{figure}[hbt]
	\includegraphics[width=\columnwidth]{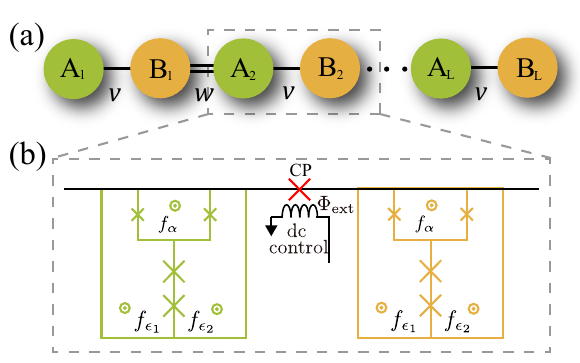}
	\caption{(a) Superconducting-qubit chain with two sites, $A_n$ and $B_n$, per unit cell and alternating couplings $v$ and $w$. (b) Schematic implementation using gap-tunable flux qubits connected by flux-controlled couplers (CPs)~\cite{Gu2017a}.}\label{fig:TopologicalQubitChain}
\end{figure}

We consider the one-dimensional array shown in Fig.~\ref{fig:TopologicalQubitChain}(a). It contains $L$ unit cells, each formed by two superconducting qubits denoted by $A_n$ and $B_n$. The exchange coupling within a cell is $v$, and the coupling from $B_n$ to $A_{n+1}$ is $w$. This notation is used throughout the paper so that a single chain is the one-segment limit of the shared-boundary construction introduced below. Both couplings can be made controllable through tunable circuit elements or parametrically activated exchange interactions~\cite{liu2006controllable,Liu2007,Grajcar2006,Rigetti2005,Bertet2006,Niskanen2006,Hime2006,Niskanen2007,Harris2007,VanDerPloeg2007,Allman2010,Bialczak2011}.

The construction is not tied to one superconducting-qubit modality. Flux qubits, transmons, xmons, and gmons all provide possible implementations~\cite{Orlando1999,Liu2005a,Koch2007,Barends2013,chen2014qubit,geller2015tunable}. For definiteness, Fig.~\ref{fig:TopologicalQubitChain}(b) illustrates gap-tunable flux qubits~\cite{Paauw2009,Paauw2009Thesis,Fedorov2010,Schwarz2013,zhu2010coherent}. In this circuit the transition frequencies and nearest-neighbor exchange couplings are separate control resources, which is the only hardware property needed in the following effective description. The corresponding control mechanisms are summarized in Appendix~\ref{sec:chain}.

In units with $\hbar=1$, the qubit Hamiltonian reads
\begin{align}
	H_{\mathrm q}={}&\frac{1}{2}\sum_{n=1}^{L}
	\left(\omega_{A,n}\sigma_{A,n}^{z}+\omega_{B,n}\sigma_{B,n}^{z}\right)
	\nonumber\\
	&+v\sum_{n=1}^{L}
	\left(\sigma_{A,n}^{+}\sigma_{B,n}^{-}+\mathrm{H.c.}\right)
	\nonumber\\
	&+w\sum_{n=1}^{L-1}
	\left(\sigma_{B,n}^{+}\sigma_{A,n+1}^{-}+\mathrm{H.c.}\right).
	\label{eq:Hamiltonian_1}
\end{align}
Here $\omega_{A,n}$ and $\omega_{B,n}$ are the two sublattice frequencies, and $\sigma_{\alpha,n}^{\pm}$ acts on qubit $\alpha_n$, with $\alpha=A,B$. Equation~(\ref{eq:Hamiltonian_1}) commutes with the excitation-number operator $\hat N_{\mathrm{ex}}=\sum_{n,\alpha}\sigma_{\alpha,n}^{+}\sigma_{\alpha,n}^{-}$. The single-excitation sector is therefore closed under the dynamics and can be used without introducing a Jordan--Wigner transformation. We denote its basis states by $\vert A_n\rangle=\sigma_{A,n}^{+}\vert\mathrm{vac}\rangle$ and $\vert B_n\rangle=\sigma_{B,n}^{+}\vert\mathrm{vac}\rangle$, where $\vert\mathrm{vac}\rangle$ is the state in which every qubit is in its ground state.

We take the sublattice frequencies to be uniform within each sublattice, $\omega_{A,n}=\omega_0+\Delta_A$ and $\omega_{B,n}=\omega_0+\Delta_B$. After removing the common energy $\omega_0$, the projection of Eq.~(\ref{eq:Hamiltonian_1}) onto the single-excitation sector is
\begin{align}
	H_{\mathrm{RM}}={}&
	\Delta_A\sum_{n=1}^{L}\vert A_n\rangle\langle A_n\vert
	+\Delta_B\sum_{n=1}^{L}\vert B_n\rangle\langle B_n\vert
	\nonumber\\
	&+v\sum_{n=1}^{L}
	\left(\vert A_n\rangle\langle B_n\vert+\mathrm{H.c.}\right)
	\nonumber\\
	&+w\sum_{n=1}^{L-1}
	\left(\vert B_n\rangle\langle A_{n+1}\vert+\mathrm{H.c.}\right).
	\label{eq:Rice-Mele}
\end{align}
For the staggered choice $\Delta_A=\Delta$ and $\Delta_B=-\Delta$, Eq.~(\ref{eq:Rice-Mele}) is the Rice--Mele Hamiltonian~\cite{Rice1982,Asboth2016}. Its SSH limit is obtained at $\Delta=0$. For positive real couplings, the open chain supports a pair of end-localized modes in the regime $\lvert v/w\rvert<1$~\cite{Asboth2016}. These modes, rather than the full $2L$-dimensional single-excitation space, provide the effective transfer channel developed in the next section.

\section{Topological analysis of edge state transfer}\label{sec:pumping}

We begin with edge-to-edge transfer in a single finite Rice--Mele chain, which provides the two-boundary-mode reference for the shared-boundary construction developed later. After presenting the microscopic control protocol and its full-chain dynamics, we derive the corresponding two-state Hamiltonian and identify the finite-size edge-mode hybridization that controls the adiabatic transfer gap.


\subsection{Pumping of an edge state}\label{pumpingA}

The Rice--Mele chain introduced in Section~\ref{sec:qubitchain} becomes a controllable edge-state channel when its intracell coupling and staggered on-site potential are varied in time~\cite{Gu2017a}. Restricting the qubit array to the conserved single-excitation sector gives the time-dependent Hamiltonian
\begin{align}
	H_{\mathrm{RM}}(s)={}&
	\Delta(s)\sum_{n=1}^{L}
	\left(\lvert A_n\rangle\langle A_n\rvert
	-\lvert B_n\rangle\langle B_n\rvert\right)
	\nonumber\\
	&+v(s)\sum_{n=1}^{L}
	\left(\lvert A_n\rangle\langle B_n\rvert+\mathrm{H.c.}\right)
	\nonumber\\
	&+w\sum_{n=1}^{L-1}
	\left(\lvert B_n\rangle\langle A_{n+1}\rvert+\mathrm{H.c.}\right),
	\label{Hamiltonian:Pumping}
\end{align}
where $s=t/T$ is the normalized time and $T$ is the duration of one control cycle. The exchange couplings and qubit frequencies required for this modulation can be implemented with the tunable circuit elements discussed in Appendix~\ref{sec:chain}.

\begin{figure}[hbt]
	\centering
 	\includegraphics[width=\columnwidth]{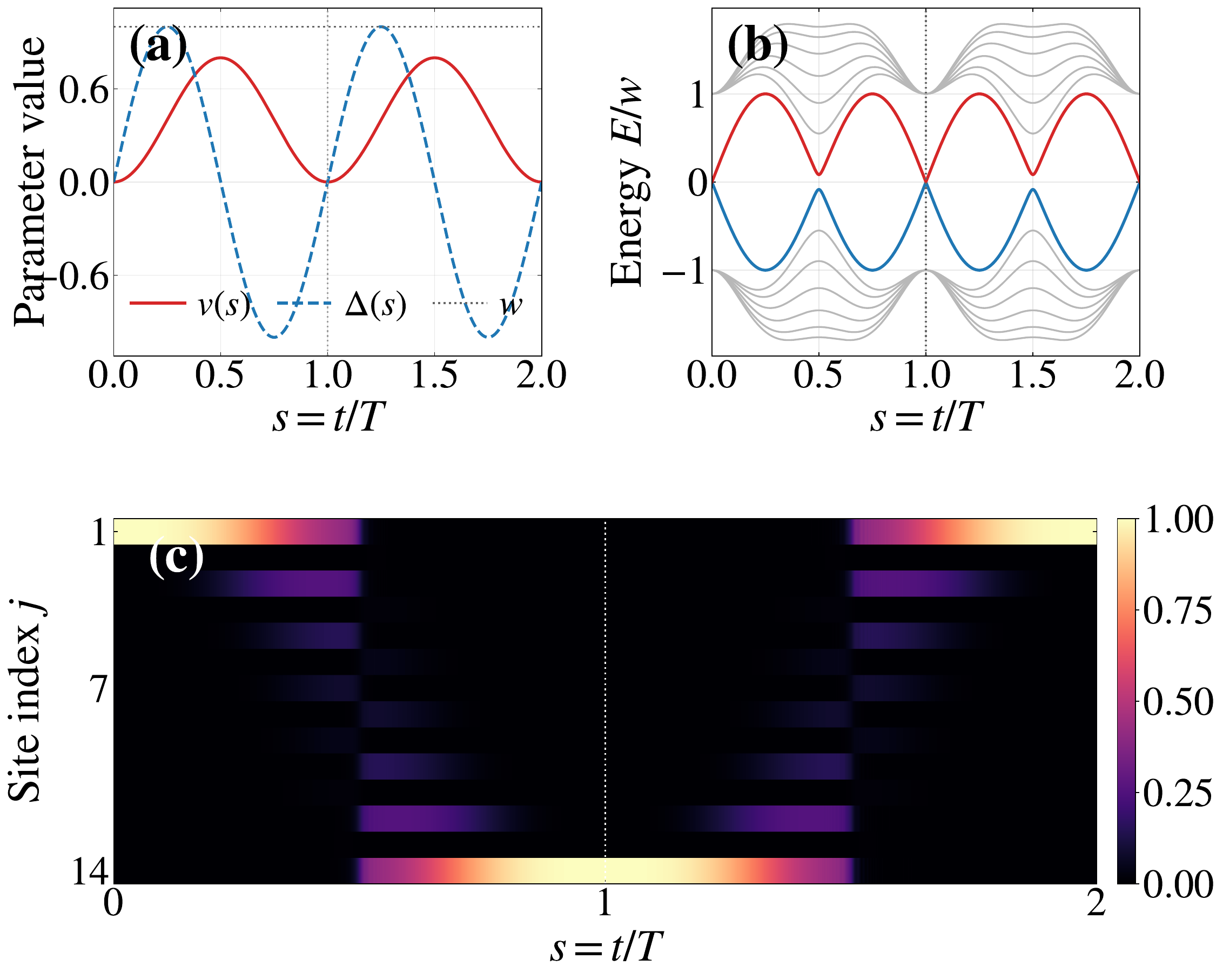}
	\caption{Two-cycle evolution of a 14-site open Rice--Mele chain. (a) Control parameters $v(s)=v_{\max}[1-\cos(2\pi s)]$ and $\Delta(s)=\Delta_{\max}\sin(2\pi s)$, with $v_{\max}=0.4$, $\Delta_{\max}=1$, and fixed $w=1$. (b) Instantaneous single-excitation spectrum; the red and blue curves highlight the two in-gap edge-state branches. (c) Site-resolved population $\lvert\psi_j(s)\rvert^2$ for an excitation initialized on site $j=1$; the color scale ranges from zero to unity. Here $s=t/T$, each cycle has duration $T=2000/w$, and the dotted lines at $s=1$ separate the two cycles. The excitation reaches the right edge after the first cycle and returns to the left edge after the second.}\label{fig:Pumping}
\end{figure}

For the two-cycle evolution in Fig.~\ref{fig:Pumping}, we use
\begin{align}
	v(s)&=v_{\max}\left[1-\cos(2\pi s)\right],\nonumber\\
	w&=1,\nonumber\\
	\Delta(s)&=\Delta_{\max}\sin(2\pi s),
	\qquad 0\leq s\leq2,
	\label{eq:pump}
\end{align}
with $v_{\max}=0.4$ and $\Delta_{\max}=1$. Thus $v(s)$ reaches $2v_{\max}=0.8<w$, while the same closed control loop is traversed once on each interval $0\leq s\leq1$ and $1\leq s\leq2$. At $s=0$, both $v$ and $\Delta$ vanish, so the first site $\lvert A_1\rangle$ is an exact left-boundary eigenstate. It can be prepared from the all-ground-state configuration by applying a $\pi$ pulse to the first qubit.

Figure~\ref{fig:Pumping}(a) shows the repeated controls, and Fig.~\ref{fig:Pumping}(b) gives the corresponding instantaneous spectrum of the 14-site open chain. We obtain the dynamics by integrating the time-dependent Schr\"odinger equation in the single-excitation sector. For $T=2000/w$, the site-resolved population in Fig.~\ref{fig:Pumping}(c) moves from the left boundary to the right boundary during the first cycle and returns to the left boundary during the second.

At the junction of the two cycles, $s=1$, the two in-gap levels cross at zero energy because $v(1)=\Delta(1)=0$. This degeneracy does not mix the two boundary states: at that instant the left and right edge sites are exactly decoupled, so the wave function remains on the boundary state reached at the end of the first cycle. The labels of the two energy branches can interchange across the crossing without producing a change of the physical state. The following subsection makes this point explicit by projecting the full chain onto its two localized boundary modes.

\subsection{Topology of the pumping process}\label{sec:TopologyAnalysis}

The transfer mechanism becomes transparent after projecting the full Rice--Mele chain onto its localized boundary modes. We first set $\Delta=0$, so that Eq.~(\ref{Hamiltonian:Pumping}) reduces to an SSH chain. For positive couplings in the regime $\lvert v/w\rvert<1$, the corresponding left and right boundary modes are
\begin{equation}
	\lvert L\rangle=\mathcal{N}\sum_{n=1}^{L}
	\lambda^{n-1}\lvert A_n\rangle
\end{equation}
and
\begin{equation}
	\lvert R\rangle=\mathcal{N}\sum_{n=1}^{L}
	\lambda^{L-n}\lvert B_n\rangle,
\end{equation}
where
$\lambda=-v/w$ and
$\mathcal{N}=\sqrt{(1-\lambda^2)/(1-\lambda^{2L})}$.
These states are exponentially localized at opposite ends and occupy different sublattices. For a finite chain they are not completely independent: their exponentially small overlap produces the matrix element that drives end-to-end transfer. When the two boundary modes remain spectrally isolated from the bulk, they form a controlled low-energy basis for the dynamics~\cite{Asboth2016,longhi2019landau,chen2021,shen2021acoustic}. Projecting Eq.~(\ref{Hamiltonian:Pumping}) onto the ordered basis $\{\lvert L\rangle,\lvert R\rangle\}$ gives
\begin{equation}\label{eq:u}
	\langle L\rvert H_{\mathrm{RM}}\lvert L\rangle
	=-\langle R\rvert H_{\mathrm{RM}}\lvert R\rangle
	=\Delta,
\end{equation}
\begin{equation}\label{eq:g}
	\langle L\rvert H_{\mathrm{RM}}\lvert R\rangle
	=\langle R\rvert H_{\mathrm{RM}}\lvert L\rangle
	=\mathcal{N}^{2}v\lambda^{L-1}\equiv g.
\end{equation}
The resulting two-boundary-mode Hamiltonian is therefore
\begin{equation}
	H_{\mathrm{eff}}^{(2)}=
	\begin{pmatrix}
		\Delta & g\\
		g & -\Delta
	\end{pmatrix},
	\label{LZmodel}
\end{equation}
which has the standard Landau--Zener structure. Here $2\Delta$ is the boundary-mode detuning and $g$ is the finite-size hybridization. The instantaneous boundary modes can be chosen real and normalized; because they occupy disjoint sublattices, their off-diagonal geometric connection vanishes, while the diagonal connection can be removed by a phase convention. Equation~(\ref{LZmodel}) therefore contains the complete projected mixing between the two modes. In particular, $g\propto(v/w)^{L-1}$ makes explicit the central localization--speed tradeoff: stronger boundary localization suppresses the coupling that opens the transfer gap. This microscopic relation between the full chain and a few-state channel is the key construction used again for the shared-boundary geometry below.

\begin{figure}[hbt]
	\centering
	\includegraphics[width=\columnwidth]{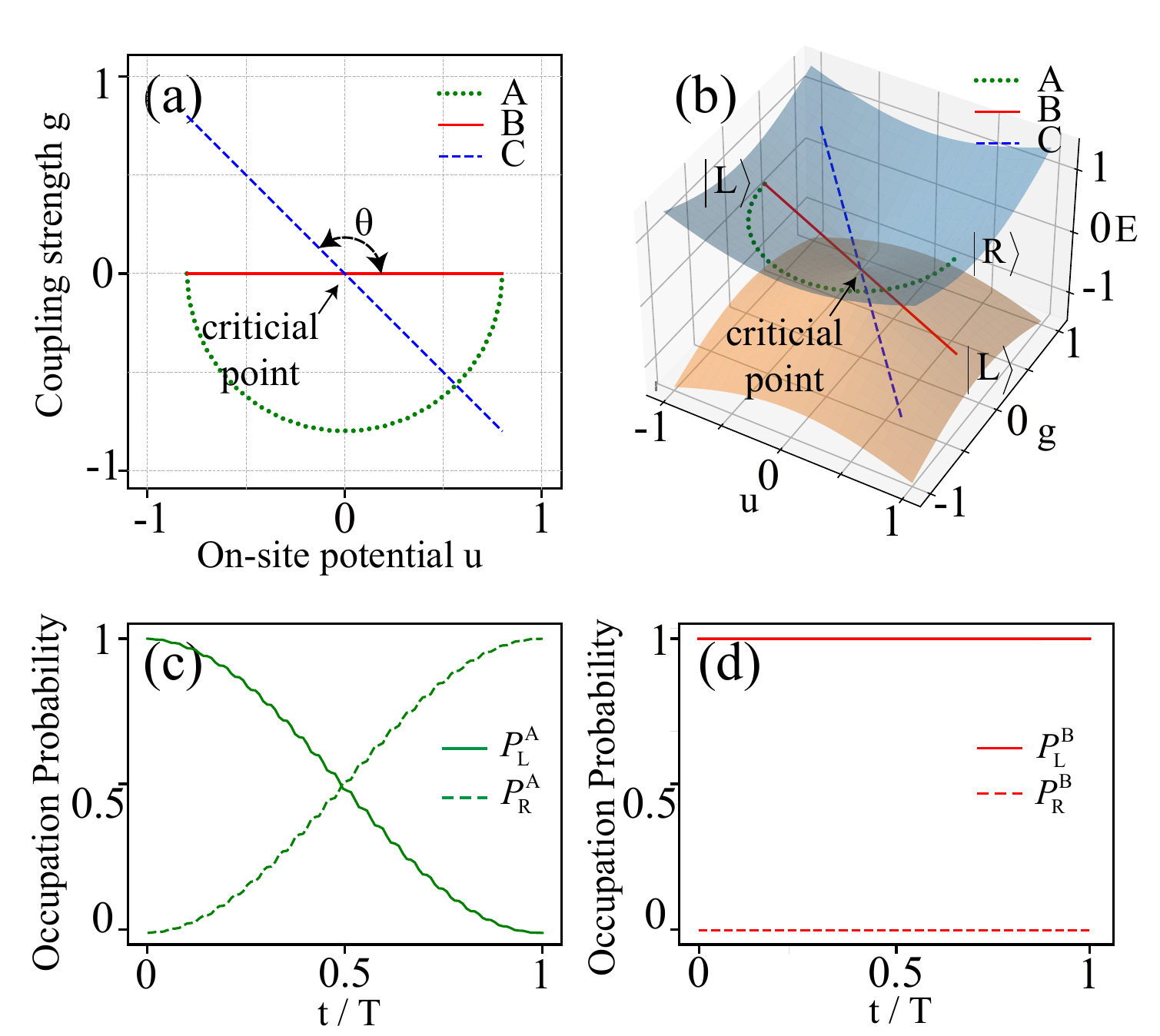}
	\caption{Boundary-mode interpretation of three representative paths. (a) Paths A, B, and C in the effective parameter plane spanned by the detuning $\Delta$ and hybridization $g$. (b) The two eigenenergy surfaces of Eq.~(\ref{LZmodel}). Path A avoids the degeneracy, whereas paths B and C pass through it. (c) Boundary-mode populations along path A. (d) Boundary-mode populations along path B.}\label{fig:LZ1}
\end{figure}

The effective Hamiltonian in Eq.~(\ref{LZmodel}) can be viewed geometrically in the $(\Delta,g)$ plane~\cite{landau2013quantum,yatsenko2002topology}. Its instantaneous eigenenergies are
\begin{equation}
	E_{\pm}=\pm\sqrt{\Delta^{2}+g^{2}}.
\end{equation}
As shown in Fig.~\ref{fig:LZ1}(b), the two surfaces form a double cone and meet only at $(\Delta,g)=(0,0)$. This representation distinguishes two physically different passages: a path that avoids the origin follows a nondegenerate instantaneous eigenstate, whereas a path through the origin reaches an exact degeneracy and must instead be interpreted from the coupling structure of the Hamiltonian.

To compare these cases without changing the endpoints, let $\tau=t/T\in[0,1]$. Path A is a semicircle in the effective parameter plane,
\begin{equation}
	\Delta=\alpha\cos(\pi\tau),
	\qquad g=\alpha\sin(\pi\tau).
\end{equation}
Its gap is constant, $E_+-E_-=2\alpha$. Consequently, sufficiently slow evolution keeps a state initialized as $\lvert L\rangle$ on the upper branch in the corresponding instantaneous eigenstate~\cite{messiah1962quantum,berry1984quantal}. As the detuning reverses sign, that eigenstate rotates continuously into $\lvert R\rangle$, producing the population transfer shown in Fig.~\ref{fig:LZ1}(c). By contrast, path B passes directly through the origin,
\begin{equation}
	\Delta=\alpha(1-2\tau),
	\qquad g=0.
\end{equation}
Here the instantaneous gap closes at $\tau=1/2$, so the usual nondegenerate adiabatic theorem does not apply at the crossing. Nevertheless, $g=0$ throughout the path, and the two localized boundary states are exactly decoupled. The state therefore retains its boundary identity even though its energy changes from one cone surface to the other, as verified in Fig.~\ref{fig:LZ1}(d).

The same conclusion extends to a straight path through the origin at a fixed angle $\theta$,
\begin{equation}
	\Delta=\alpha(1-2\tau),
	\qquad g=\tan(\theta)\Delta,
\end{equation}	
shown as path C in Fig.~\ref{fig:LZ1}(a). Along this path, Eq.~(\ref{LZmodel}) becomes
\begin{equation}
	H_{\mathrm{eff}}^{(2)}=
	\frac{\Delta}{\cos\theta}\begin{pmatrix}
		\cos\theta & \sin\theta\\
		\sin\theta & -\cos\theta
	\end{pmatrix}.
	\label{Hamiltonian C}
\end{equation}
Its eigenstates can be chosen as the time-independent combinations
\begin{equation}
	\begin{pmatrix}
		\lvert\uparrow\rangle\\
		\lvert\downarrow\rangle
	\end{pmatrix}
	=\begin{pmatrix}
		\cos\frac{\theta}{2} & \sin\frac{\theta}{2}\\
		\sin\frac{\theta}{2} & -\cos\frac{\theta}{2}
	\end{pmatrix}
	\begin{pmatrix}
		\lvert L\rangle\\
		\lvert R\rangle
	\end{pmatrix}.
\end{equation}
The corresponding eigenvalues are $\pm\Delta/\cos\theta$. In this fixed basis the Hamiltonian is diagonal,
\begin{equation}
	H_{\mathrm{eff}}^{(2)}=
	\frac{\Delta}{\cos\theta}\begin{pmatrix}
		1 & 0\\
		0 & -1
	\end{pmatrix}.
\end{equation}
Thus neither component can drive a transition into the other. A state prepared in $\lvert\uparrow\rangle$ or $\lvert\downarrow\rangle$ remains in that same vector while its eigenenergy label changes sign at the origin. Path B is the special case $\theta=0$, for which the fixed eigenvectors are simply $\lvert L\rangle$ and $\lvert R\rangle$. The essential distinction is therefore not whether an energy curve appears to cross, but whether the projected Hamiltonian contains a finite matrix element that rotates one localized boundary mode into the other.

\begin{figure}[hbt]
	\centering
	\includegraphics[width=\columnwidth]{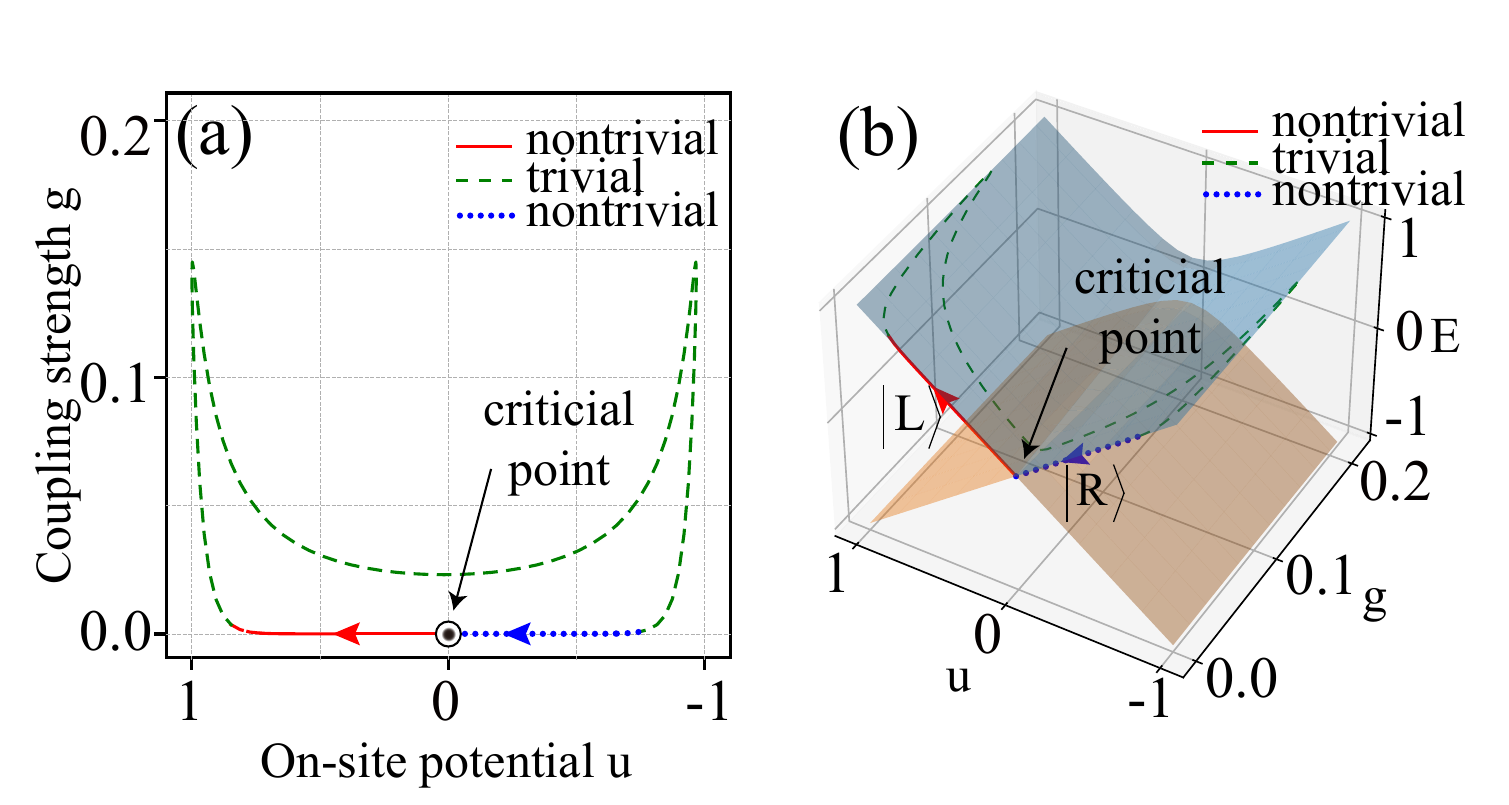}
	\caption{Schematic boundary-mode representation of the Rice--Mele control sequence. (a) Effective trajectory in the parameter plane spanned by $\Delta$ and $g$. The line styles distinguish successive portions of the microscopic evolution. (b) The same trajectory on the eigenenergy surfaces of Eq.~(\ref{LZmodel}). The origin is the exact crossing at which the two localized boundary modes decouple.}\label{fig:LZ2}
\end{figure}

For the microscopic protocol in Section~\ref{pumpingA}, Eqs.~(\ref{eq:pump}), (\ref{eq:u}), and (\ref{eq:g}) determine the effective path without fitting parameters. Figure~\ref{fig:LZ2}(a) illustrates how the microscopic controls are converted into a trajectory in the $(\Delta,g)$ plane, while Fig.~\ref{fig:LZ2}(b) lifts that trajectory onto the two eigenenergy surfaces. During the first cycle, the state starts as $\lvert L\rangle$, follows the upper boundary-mode branch through the finite avoided crossing at $\Delta=0$ and $g\neq0$, and ends as $\lvert R\rangle$. At $s=1$, both $\Delta$ and $g$ vanish. The trajectory therefore reaches the conical intersection, but the state itself is unchanged because the two boundary modes are then decoupled. The second cycle traverses the same control path from the state $\lvert R\rangle$ and follows the lower branch back to $\lvert L\rangle$.

The projected model identifies the boundary-mode detuning, finite-size coupling, and crossing structure, while the full-chain spectrum determines whether the two modes remain isolated from the bulk.

\section{Topological edge-state transfer through a shared boundary}\label{sec:shared_boundary}

The two-state reduction discussed above describes the hybridization of the left and right edge states of a single Rice--Mele chain. We now extend this construction by joining two finite Rice--Mele subchains at one physical boundary qubit. The shared qubit provides an additional localized boundary degree of freedom, so that the relevant low-energy subspace becomes three dimensional. We retain independent intracell and intercell couplings $v_1,w_1$ and $v_2,w_2$ for the two subchains, as well as independent on-site potentials $\Delta_1,\Delta_2,\Delta_3$ for the three boundary-mode sectors.

\subsection{Shared-boundary Rice--Mele chain}

We label the sublattices of the two subchains by $A_{\mu,n}$ and $B_{\mu,n}$, where $\mu=1,2$ and $n=1,\ldots,L$. The right boundary qubit of the first subchain and the left boundary qubit of the second subchain are identified as the same physical site,
\begin{equation}
	\vert S\rangle\equiv\vert B_{1,L}\rangle=\vert A_{2,1}\rangle .
\end{equation}
This identification is physical rather than an additional coupling between two distinct qubits. Consequently, the composite chain contains $4L-1$ physical sites, and the state $\vert S\rangle$ must be counted only once.

Starting from the Rice--Mele model in Eq.~(\ref{eq:Rice-Mele}), the hopping part of the shared-boundary Hamiltonian is
\begin{align}
	H_{\mathrm{hop}}={}&
	v_1\sum_{n=1}^{L}
	\left(\vert A_{1,n}\rangle\langle B_{1,n}\vert+\mathrm{H.c.}\right)
	\nonumber\\
	&+w_1\sum_{n=1}^{L-1}
	\left(\vert B_{1,n}\rangle\langle A_{1,n+1}\vert+\mathrm{H.c.}\right)
	\nonumber\\
	&+v_2\sum_{n=1}^{L}
	\left(\vert A_{2,n}\rangle\langle B_{2,n}\vert+\mathrm{H.c.}\right)
	\nonumber\\
	&+w_2\sum_{n=1}^{L-1}
	\left(\vert B_{2,n}\rangle\langle A_{2,n+1}\vert+\mathrm{H.c.}\right).
	\label{eq:shared_hopping}
\end{align}
The three on-site potentials are applied to the sublattice sectors supporting the left, shared, and right boundary modes, respectively:
\begin{align}
	H_{\Delta}={}&
	\Delta_1\sum_{n=1}^{L}
	\vert A_{1,n}\rangle\langle A_{1,n}\vert
	\nonumber\\
	&+\Delta_2\sum_{n=1}^{L}
	\vert B_{1,n}\rangle\langle B_{1,n}\vert
	\nonumber\\
	&+\Delta_2\sum_{n=1}^{L}
	\vert A_{2,n}\rangle\langle A_{2,n}\vert
	-\Delta_2\vert S\rangle\langle S\vert
	\nonumber\\
	&+\Delta_3\sum_{n=1}^{L}
	\vert B_{2,n}\rangle\langle B_{2,n}\vert .
	\label{eq:shared_onsite}
\end{align}
The subtraction in the $\Delta_2$ term removes the duplicate contribution of the shared physical qubit. The complete microscopic Hamiltonian is
\begin{equation}
	H_{\mathrm{SB}}=H_{\mathrm{hop}}+H_{\Delta}.
	\label{eq:shared_micro}
\end{equation}

We focus on the edge-state regime
\begin{equation}
	\left\vert\frac{v_1}{w_1}\right\vert<1,\qquad
	\left\vert\frac{v_2}{w_2}\right\vert<1,
\end{equation}
and define the two decay factors
\begin{equation}
	\lambda_1=-\frac{v_1}{w_1},\qquad
	\lambda_2=-\frac{v_2}{w_2}.
\end{equation}
For later use, we introduce
\begin{equation}
	G_L(\lambda_i)=\sum_{m=0}^{L-1}\vert\lambda_i\vert^{2m}
	=\frac{1-\vert\lambda_i\vert^{2L}}{1-\vert\lambda_i\vert^2},
	\qquad i=1,2.
\end{equation}
The normalization factors are
\begin{align}
	\mathcal{N}_1&=G_L^{-1/2}(\lambda_1),\nonumber\\
	\mathcal{N}_2&=
	\left[G_L(\lambda_1)+G_L(\lambda_2)-1\right]^{-1/2},
	\nonumber\\
	\mathcal{N}_3&=G_L^{-1/2}(\lambda_2).
	\label{eq:shared_normalization}
\end{align}
The $-1$ in $\mathcal{N}_2$ again removes the double counting of $\vert S\rangle$.

The three localized boundary states can then be chosen as
\begin{align}
	\vert E_1\rangle={}&
	\mathcal{N}_1\sum_{n=1}^{L}\lambda_1^{n-1}\vert A_{1,n}\rangle ,
	\nonumber\\
	\vert E_2\rangle={}&
	\mathcal{N}_2\left(
	\vert S\rangle
	+\sum_{n=1}^{L-1}\lambda_1^{L-n}\vert B_{1,n}\rangle
	+\sum_{n=2}^{L}\lambda_2^{n-1}\vert A_{2,n}\rangle
	\right),
	\nonumber\\
	\vert E_3\rangle={}&
	\mathcal{N}_3\sum_{n=1}^{L}\lambda_2^{L-n}\vert B_{2,n}\rangle .
	\label{eq:shared_modes}
\end{align}
Here $\vert E_1\rangle$ and $\vert E_3\rangle$ are localized at the two outer ends, whereas $\vert E_2\rangle$ is centered on the shared physical qubit and has generally different exponential tails in the two subchains. Since the three states occupy disjoint sublattice sets, they satisfy $\langle E_i\vert E_j\rangle=\delta_{ij}$. For a finite chain they form a localized low-energy basis rather than three exact eigenstates; their finite-size hybridization produces the effective couplings below.

\subsection{Effective Hamiltonian in the boundary-state subspace}

The first subchain determines the coupling between the left and shared boundary modes. Direct evaluation gives
\begin{align}
	g_1={}&\langle E_1\vert H_{\mathrm{SB}}\vert E_2\rangle
	\nonumber\\
	={}&\mathcal{N}_1\mathcal{N}_2
	\left[
	Lv_1\lambda_1^{L-1}+(L-1)w_1\lambda_1^L
	\right]
	\nonumber\\
	={}&\mathcal{N}_1\mathcal{N}_2v_1\lambda_1^{L-1},
	\label{eq:shared_g1}
\end{align}
where $w_1\lambda_1=-v_1$ has been used. Similarly, the second subchain gives
\begin{align}
	g_2={}&\langle E_2\vert H_{\mathrm{SB}}\vert E_3\rangle
	\nonumber\\
	={}&\mathcal{N}_2\mathcal{N}_3
	\left[
	Lv_2\lambda_2^{L-1}+(L-1)w_2\lambda_2^L
	\right]
	\nonumber\\
	={}&\mathcal{N}_2\mathcal{N}_3v_2\lambda_2^{L-1},
	\label{eq:shared_g2}
\end{align}
where $w_2\lambda_2=-v_2$. Thus $\vert g_i\vert$ scales as $\vert v_i/w_i\vert^L$ for the corresponding subchain. Because the microscopic Hamiltonian contains only nearest-neighbor hopping and the two outer modes belong to different subchains,
\begin{equation}
	\langle E_1\vert H_{\mathrm{SB}}\vert E_3\rangle=0.
\end{equation}

The on-site potential in each sector is uniform over the support of its boundary mode. Its projection is therefore
\begin{align}
	\langle E_1\vert H_{\mathrm{SB}}\vert E_1\rangle
	&=\Delta_1,\nonumber\\
	\langle E_2\vert H_{\mathrm{SB}}\vert E_2\rangle
	&=\Delta_2,\nonumber\\
	\langle E_3\vert H_{\mathrm{SB}}\vert E_3\rangle
	&=\Delta_3.
\end{align}
Introducing the projector
\begin{equation}
	P_{\mathrm{edge}}=\sum_{j=1}^{3}\vert E_j\rangle\langle E_j\vert ,
\end{equation}
the microscopic Hamiltonian reduces in the ordered basis
$\{\vert E_1\rangle,\vert E_2\rangle,\vert E_3\rangle\}$ to
\begin{equation}
	H_{\mathrm{eff}}^{(3)}
	=P_{\mathrm{edge}}H_{\mathrm{SB}}P_{\mathrm{edge}}
	=
	\begin{pmatrix}
		\Delta_1 & g_1 & 0\\
		g_1 & \Delta_2 & g_2\\
		0 & g_2 & \Delta_3
	\end{pmatrix}.
	\label{eq:shared_eff_general}
\end{equation}
This projection reduces the full $(4L-1)$-dimensional single-excitation problem to a general three-node channel formed by the left edge mode, the shared boundary mode, and the right edge mode. Its instantaneous eigenenergies are the roots of
\begin{align}
	0={}&
	(E-\Delta_1)(E-\Delta_2)(E-\Delta_3)
	\nonumber\\
	&-g_1^2(E-\Delta_3)-g_2^2(E-\Delta_1).
	\label{eq:shared_characteristic}
\end{align}
Equations~(\ref{eq:shared_g1}), (\ref{eq:shared_g2}), and (\ref{eq:shared_eff_general}) retain the independent microscopic parameters $v_1,v_2,w_1,w_2$ and the three boundary-mode potentials $\Delta_1,\Delta_2,\Delta_3$. The two protocols below correspond to different time-dependent choices of these effective parameters.

\begin{figure}[t]
	\centering
	\includegraphics[width=\columnwidth]{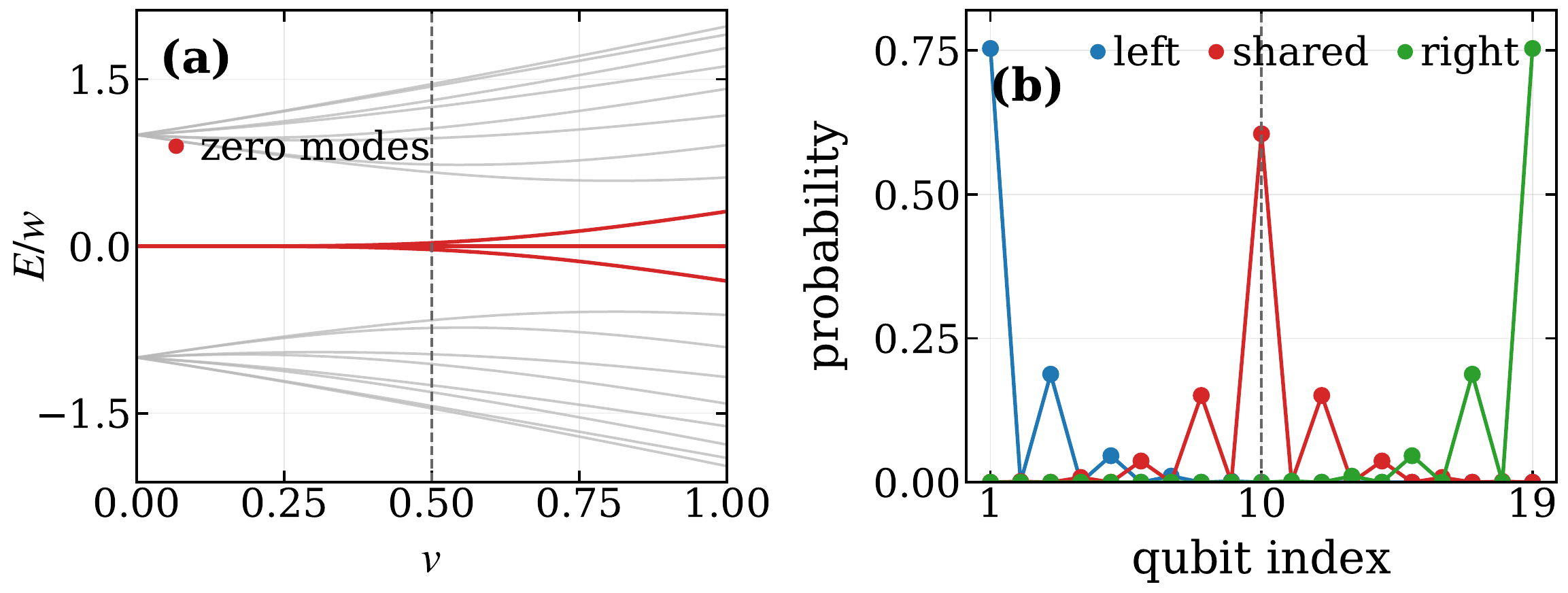}
	\caption{Full spectrum and localized boundary modes of two SSH subchains sharing one endpoint. (a) Spectrum of the full 19-site Hamiltonian for $w_1=w_2=1$ and $v_1=v_2=v$. The three central in-gap branches are highlighted in red, and the vertical dashed line indicates $v=0.5$. (b) Site-resolved probabilities of the three orthogonal localized modes obtained by diagonalizing the projected position operator in the three-dimensional near-zero-energy spectral subspace at $v=0.5$.}
	\label{fig:shared_three_modes}
\end{figure}

To compare this boundary-state projection directly with the full microscopic spectrum, we now consider the symmetric parameter cut $w_1=w_2=1$, $v_1=v_2=v$, and $0\leq v\leq1$. We take $L=5$, so that each SSH subchain contains $2L=10$ qubits and the identification of one common endpoint leaves $4L-1=19$ physical qubits. Direct diagonalization of the full 19-dimensional single-excitation Hamiltonian gives the spectrum in Fig.~\ref{fig:shared_three_modes}(a). All levels are shown in gray, while the three branches closest to zero energy are highlighted in red. The vertical dashed line marks $v=0.5$. Because the chains are finite, these three boundary modes are not all exact zero-energy eigenstates: they consist of one strict zero mode and a pair of near-zero modes split by finite-size hybridization.

The three energy eigenstates nearest zero are generally delocalized linear combinations of the boundary degrees of freedom. To recover a localized basis, let $P_0$ project onto the spectral subspace spanned by these three states and define the position operator $X=\sum_{j=1}^{4L-1}j\vert j\rangle\langle j\vert$. Diagonalizing the projected operator $P_0XP_0$ within this subspace produces three mutually orthogonal combinations localized at the left endpoint, the shared boundary, and the right endpoint. Their site-resolved probabilities at $v=0.5$ are plotted in Fig.~\ref{fig:shared_three_modes}(b), where the lines connect neighboring physical sites and the markers denote the actual numerical values. These three localized combinations are precisely the microscopic representatives of $\vert E_1\rangle$, $\vert E_2\rangle$, and $\vert E_3\rangle$ used in the effective Hamiltonian above.

\subsection{Dark-state adiabatic transfer in the three-node channel}

The coupling structure of Eq.~(\ref{eq:shared_eff_general}) is identical to that of a $\Lambda$-type three-level system. For the dark-state protocol we set $\Delta_1=\Delta_2=\Delta_3=0$. The three instantaneous eigenenergies are then
\begin{equation}
	E_D=0,\qquad E_{\pm}=\pm\Omega(t),\qquad
	\Omega(t)=\sqrt{g_1^2(t)+g_2^2(t)}.
	\label{eq:shared_dark_spectrum}
\end{equation}
The zero-energy eigenstate contains no component of the shared boundary mode $\vert E_2\rangle$ and is given by
\begin{equation}
	\vert D(t)\rangle=
	\frac{g_2(t)\vert E_1\rangle-g_1(t)\vert E_3\rangle}
	{\sqrt{g_1^2(t)+g_2^2(t)}}.
	\label{eq:shared_dark_state}
\end{equation}
This form suggests the counterintuitive coupling sequence familiar from adiabatic passage. If $g_1$ is initially closed while $g_2$ is open, the dark state coincides with $\vert E_1\rangle$; slowly opening $g_1$ while closing $g_2$ rotates it continuously into $\vert E_3\rangle$. When the modulation is slow compared with the dark-to-bright-state gap $\Omega(t)$, the system follows this channel while ideally avoiding population of $\vert E_2\rangle$.

In the microscopic SSH chains, $g_1(t)$ and $g_2(t)$ are not independent controls but are generated by the weak bonds $v_1(t)$ and $v_2(t)$. Defining $s=t/T$ and fixing $w_1=w_2=w=1$, we use
\begin{align}
	f(s)&=\frac{1}{2}\sin^2(\pi s),\\
	v_1(s)&=v_{\max}\left[1-\cos(\pi s)+f(s)\right],\\
	v_2(s)&=v_{\max}\left[1+\cos(\pi s)+f(s)\right],
	\label{eq:shared_dark_pulses}
\end{align}
with $v_{\max}=0.4$. The common term $v_{\max}f(s)$ vanishes at both endpoints and raises both weak bonds from $0.4$ to $0.6$ at the midpoint. It therefore increases the central gap without changing the endpoint values $v_1(0)=0$, $v_2(0)=0.8$ and their interchange at $s=1$. Throughout the protocol, $0\leq v_i\leq0.8<w$, so the two subchains remain in the edge-state-localized regime.

For this ordering, $g_1$ is initially closed and $g_2$ is open. The initial dark state is therefore $\vert E_1\rangle$, and the protocol transports the excitation from left to right. At each instant, the effective couplings are obtained from the microscopic matrix elements derived above,
\begin{align}
	g_1(s)&=\mathcal N_1(s)\mathcal N_2(s)v_1(s)
	\left[-\frac{v_1(s)}{w}\right]^{L-1},\\
	g_2(s)&=\mathcal N_2(s)\mathcal N_3(s)v_2(s)
	\left[-\frac{v_2(s)}{w}\right]^{L-1}.
	\label{eq:shared_dark_effective_couplings}
\end{align}
The normalization factors vary together with the weak bonds. For $L=5$, $L-1=4$ is even, and both effective couplings are positive along the chosen path. The minimum effective dark-to-bright-state gap is $\min_s\Omega(s)=0.06080$.

We numerically evolve the three-state equation
\begin{equation}
	i\frac{\partial}{\partial t}\boldsymbol{c}(t)
	=H_{\mathrm{eff}}^{(3)}(t)\boldsymbol{c}(t)
	\label{eq:shared_dark_evolution}
\end{equation}
for a total duration $T=800$, starting from $c_1(0)=1$. The instantaneous localized basis states can be chosen real and normalized, and they occupy mutually disjoint sublattice sectors. Their off-diagonal geometric connections therefore vanish, so no additional basis-motion term is required in Eq.~(\ref{eq:shared_dark_evolution}). We then use the same $v_1(t)$ and $v_2(t)$ to evolve the full 19-qubit Hamiltonian from the leftmost bare state $\vert1\rangle$.

\begin{figure}[t]
	\centering
	\includegraphics[width=\columnwidth]{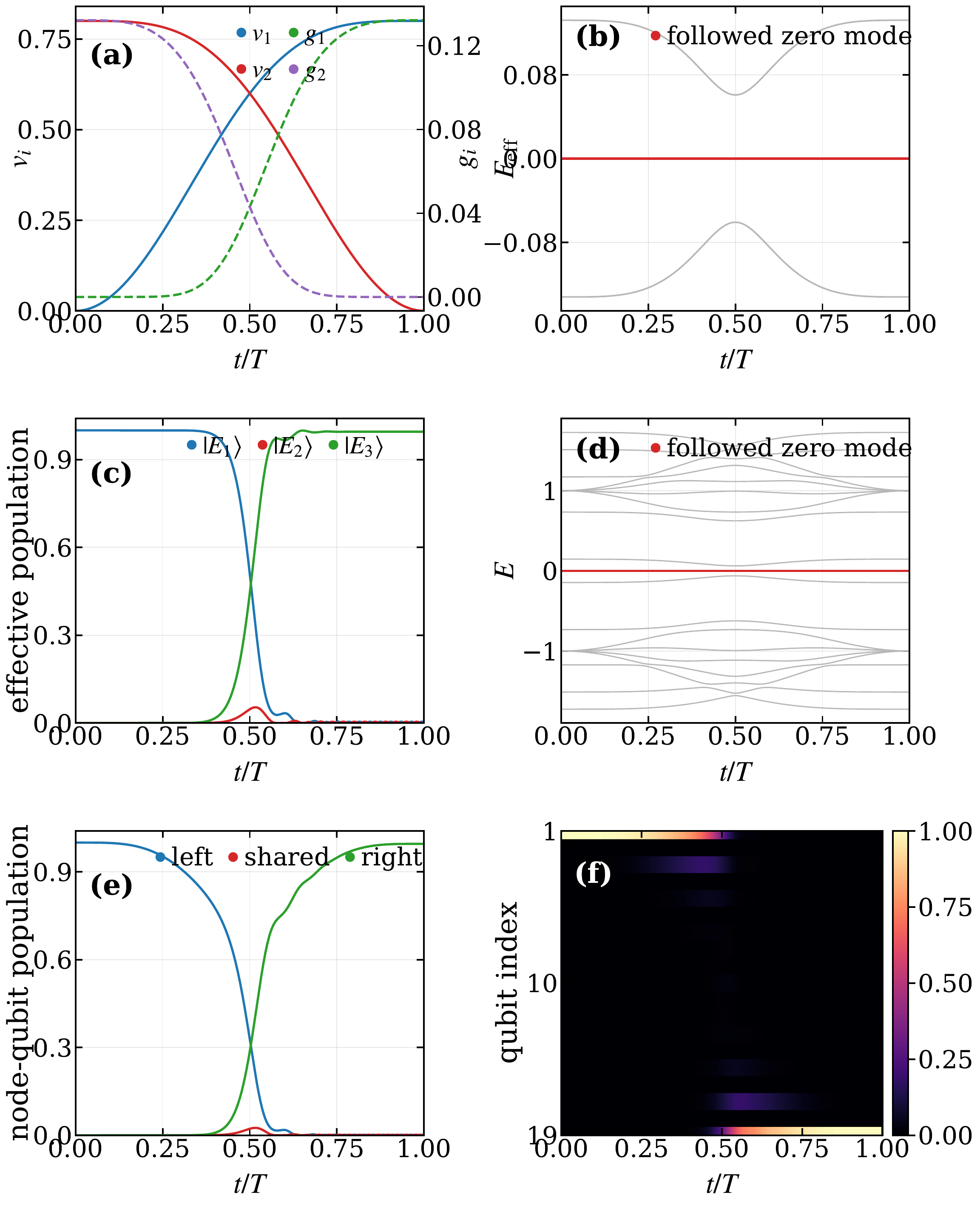}
	\caption{Dark-state adiabatic transfer through the shared-boundary three-node channel. (a) Microscopic weak bonds $v_1,v_2$ and the corresponding effective couplings $g_1,g_2$. (b) Instantaneous spectrum of the three-state Hamiltonian, with the followed zero-energy dark state highlighted. (c) Effective-mode populations. (d) Spectrum of the full 19-qubit chain, with the corresponding zero-energy branch highlighted. (e) Populations of the left, shared, and right physical boundary qubits. (f) Population evolution from qubit 1 to qubit 19 over the full chain, with qubit 1 at the top and qubit 19 at the bottom. The parameters are $L=5$, $v_{\max}=0.4$, and $T=800$.}
	\label{fig:shared_dark_transfer}
\end{figure}

Figure~\ref{fig:shared_dark_transfer} compares the effective and microscopic dynamics. The minimum separation of the followed zero-energy branch from its neighboring levels is $0.06080$ in the effective model and $0.06184$ in the full chain. The final right-mode population in the three-state model is $P_{E_3}(T)=0.99530$, while the final population of the rightmost qubit in the full chain is $P_{19}(T)=0.99569$. The maximum population of $\vert E_2\rangle$ in the effective model is $0.0539$, whereas the maximum population of the physical shared qubit is only $0.0254$. Thus, the enhanced-midpoint weak-bond path maintains high-fidelity end-to-end transfer for the $L=5$ subchains. The process is primarily a continuous exchange of weight between the two outer boundary modes, rather than a highly occupied wave packet propagating sequentially through every bulk site.

\subsection{Continuous Rice--Mele-type adiabatic transfer}

The general three-state Hamiltonian also supports transport along a nonzero-energy instantaneous eigenbranch when time-dependent on-site potentials are introduced. This open Rice--Mele-type protocol complements transfer through a zero-energy dark state. We consider two subchains of length $L=5$, so that the shared-boundary chain contains $4L-1=19$ physical qubits. With the normalized time $s=t/T$, we choose
\begin{align}
	w_1(s)&=w_2(s)=1,\\
	v_1(s)&=v_2(s)=v_{\max}\left[1-\cos(2\pi s)\right],
	\label{eq:shared_pump_couplings}
\end{align}
and impose the boundary-mode potentials
\begin{align}
	\Delta_1(s)&=-\Delta_{\max}\cos(\pi s),\\
	\Delta_2(s)&=0,\\
	\Delta_3(s)&=\Delta_{\max}\cos(\pi s).
	\label{eq:shared_pump_potentials}
\end{align}
Here $v_{\max}=0.4$ and $\Delta_{\max}=0.5$. Thus, the weak bonds vanish at the beginning and end of the protocol and reach a peak value $2v_{\max}=0.8$ at $s=1/2$. Initially, $(\Delta_1,\Delta_2,\Delta_3)=(-0.5,0,0.5)$, and the isolated state $\vert E_1\rangle$ lies on the lowest boundary-mode branch. At the end, the potentials are reversed to $(0.5,0,-0.5)$, and the same instantaneous branch becomes $\vert E_3\rangle$. The simultaneous opening of the two weak bonds converts the two sequential edge transfers into a single continuous adiabatic passage through the shared boundary.

For the symmetric path in Eqs.~(\ref{eq:shared_pump_couplings}) and (\ref{eq:shared_pump_potentials}), one has $g_1=g_2=g$ and $\Delta_1=-\Delta_3$. The eigenenergies of Eq.~(\ref{eq:shared_eff_general}) therefore reduce to
\begin{equation}
	E_0=0,\qquad E_{\pm}=\pm\sqrt{\Delta_1^2+2g^2}.
	\label{eq:shared_pump_spectrum}
\end{equation}
The transported state follows the lower branch $E_-$ rather than the zero-energy dark state. Consequently, the shared boundary mode acquires a finite transient population when the outer potentials cross at $s=1/2$; in the ideal three-state model its maximum population is $1/2$. Throughout the protocol, $v_{1,2}/w_{1,2}\leq0.8<1$, so both subchains remain in the edge-state-localized regime while their finite-size overlap opens an avoided crossing.

\begin{figure}[t]
	\centering
	\includegraphics[width=\columnwidth]{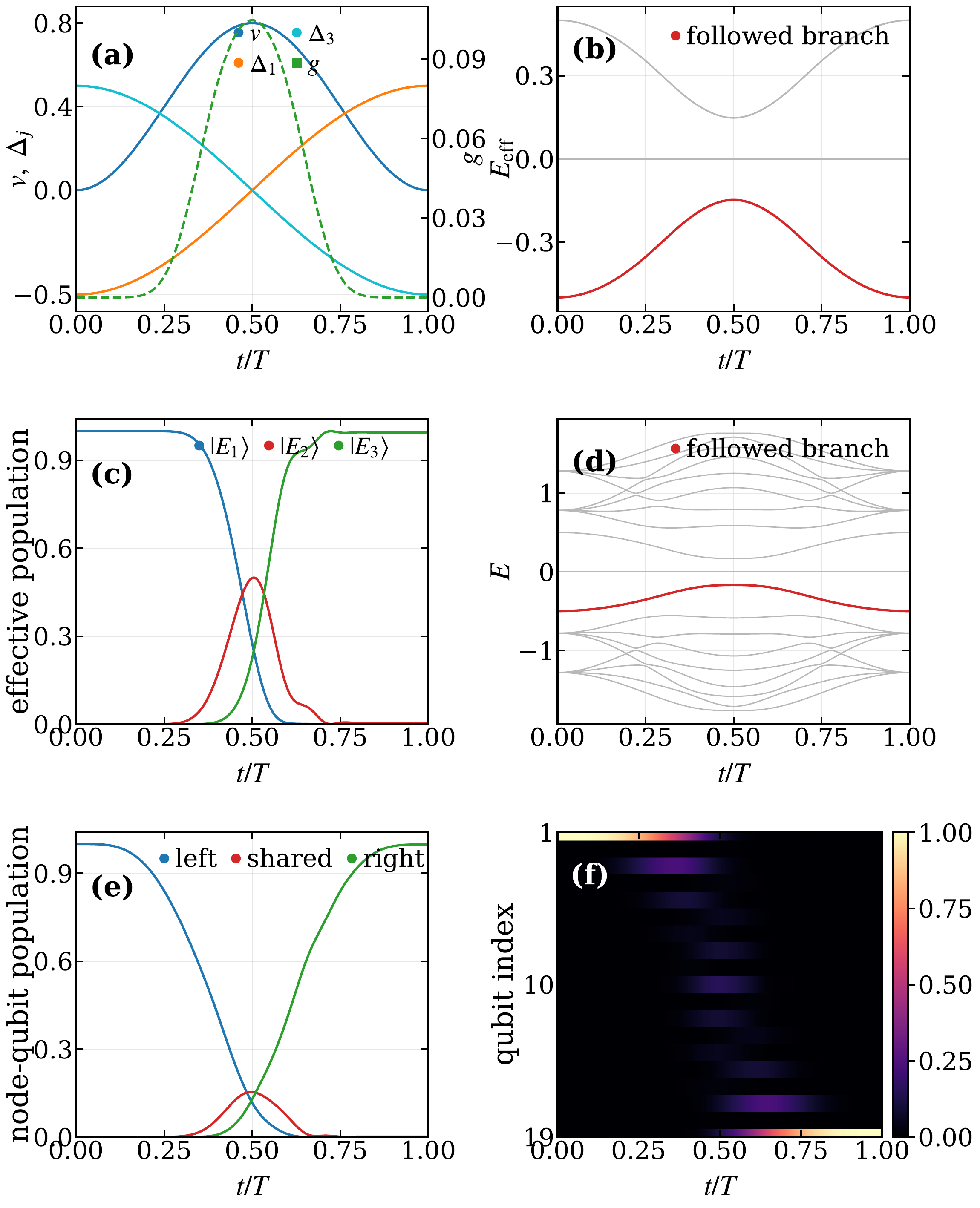}
	\caption{Continuous Rice--Mele-type adiabatic transfer through a shared boundary. (a) Weak bond $v$, effective coupling $g$, and boundary-mode potentials $\Delta_1$ and $\Delta_3$. (b) Instantaneous spectrum of the three-state effective Hamiltonian; the followed lower branch is highlighted in red. (c) Populations of the three effective boundary modes. (d) Instantaneous spectrum of the full 19-qubit chain, with the followed in-gap branch highlighted. (e) Populations of the left, shared, and right physical boundary qubits. (f) Population evolution from qubit 1 to qubit 19 over the full chain, with qubit 1 at the top and qubit 19 at the bottom. The parameters are $L=5$, $v_{\max}=0.4$, $\Delta_{\max}=0.5$, and $T=200$.}
	\label{fig:shared_boundary_pump}
\end{figure}

As shown in Fig.~\ref{fig:shared_boundary_pump}, the minimum separation of the followed branch from the other levels is $0.14791$ in the effective model and $0.16723$ in the full chain. For $T=200$, the final population of $\vert E_3\rangle$ in the effective model is $0.99535$, while the population of the rightmost qubit in the full chain reaches $0.99831$. The maximum shared-node population in the full chain is $0.1537$, smaller than the effective-mode value because $\vert E_2\rangle$ also contains exponential tails away from the shared physical qubit.

This protocol is an open Rice--Mele-type adiabatic path: it uses a reversal of the on-site potentials together with time-dependent bonds to connect two edge states, but its parameter trajectory is not a closed cycle. Therefore, the single transfer shown here is not a Chern-number-quantized Thouless pump~\cite{Thouless1983}.

\subsection{Chain-length comparison with the original single-chain transfer}

We finally compare the shared-boundary Rice--Mele transfer with the original open Rice--Mele transfer in a single chain. Before comparing their length dependence, the two protocols were optimized within the same subcritical control family. The best-performing case for both protocols is obtained with $v_{\max}=0.45$ and $\Delta_{\max}=0.25$, which we use throughout this comparison. The dark-state protocol is not included: although it can produce high-fidelity transfer for the short $L=5$ example above, its effective gap becomes too small as the segment length increases, resulting in poor transfer and substantially longer adiabatic times at the resource scales considered here. It is consequently not competitive with the two on-site-potential-driven protocols in this length comparison.

The original protocol is evaluated for single chains with $N_{\mathrm{original}}=38,58,78$, whereas the shared-boundary systems consist of two even-length segments of $20$, $30$, and $40$ sites. Identifying one common endpoint gives $N_{\mathrm{shared}}=39,59,79$. Thus, the pairs $38/39$, $58/59$, and $78/79$ have nearly identical physical resources. Both protocols use $w=1$ and
\begin{equation}
	v(s)=0.45\left[1-\cos(2\pi s)\right],\qquad s=t/T,
	\label{eq:comparison_common_bond}
\end{equation}
so that the peak weak bond is $0.9w$ and neither path reaches the uniform-chain critical point $v=w$. For the original chain, the two sublattice potentials are
\begin{equation}
	\Delta_A(s)=-0.25\cos(\pi s),\qquad
	\Delta_B(s)=0.25\cos(\pi s),
\end{equation}
while the three boundary-mode sectors of the shared system use
\begin{align}
	\Delta_1(s)&=-0.25\cos(\pi s),\\
	\Delta_2(s)&=0,\\
	\Delta_3(s)&=0.25\cos(\pi s).
\end{align}
Both evolutions start from the leftmost bare excitation with energy $-0.25$ and end on the rightmost state with the same energy. We denote the target-site population by $P_{\mathrm{tar}}$.

\begin{figure}[t]
	\centering
	\includegraphics[width=\columnwidth]{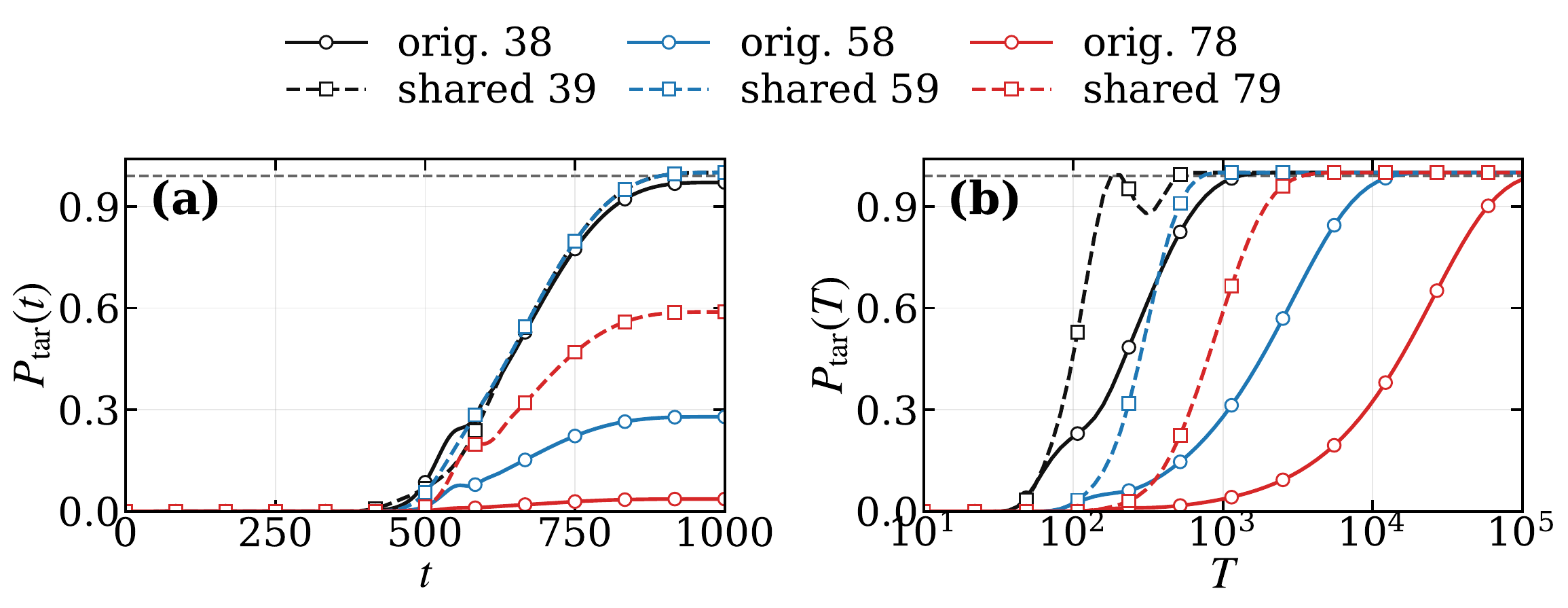}
	\caption{Length and time comparison between the original single-chain Rice--Mele transfer and the shared-boundary Rice--Mele transfer. (a) Target-site population during an evolution with fixed total time $T=1000/w$. (b) Final target-site population as a function of the total evolution time. Solid curves denote the original chains and dashed curves denote the shared-boundary chains; the gray dashed line marks $P_{\mathrm{tar}}=0.99$. Both protocols use their optimal common parameters $v_{\max}=0.45$ and $\Delta_{\max}=0.25$.}
	\label{fig:shared_vs_original_length}
\end{figure}

Figure~\ref{fig:shared_vs_original_length}(a) compares the six transfer trajectories at $T=1000/w$. The numerical results for the three nearly equal-size pairs are summarized in Table~\ref{tab:shared_length_comparison}.
\begin{table}[t]
	\caption{Final target populations at $T=1000/w$ and the time $T_{0.99}^{\mathrm{st}}$ after which all subsequent sampled points remain above $0.99$. Each entry in the final column lists the original/shared result.}
	\label{tab:shared_length_comparison}
	\begin{ruledtabular}
		\begin{tabular}{cccc}
		Pair & $P_{\mathrm{orig}}$ & $P_{\mathrm{shared}}$ & $T_{0.99}^{\mathrm{st}}$\\
		\hline
		$38/39$ & $0.97087$ & $0.999999$ & $1.30\times10^3/5.18\times10^2$\\
		$58/59$ & $0.27894$ & $0.999999$ & $1.39\times10^4/7.69\times10^2$\\
		$78/79$ & $0.03641$ & $0.58890$ & $>10^5/3.73\times10^3$\\
		\end{tabular}
	\end{ruledtabular}
\end{table}
At this fixed time, the 39- and 59-site shared systems have already reached nearly complete transfer, whereas the corresponding original chains reach only $0.9709$ and $0.2789$. The longest shared system is not yet fully adiabatic, but its population $0.5889$ remains much larger than the value $0.0364$ of the original 78-site chain.

Figure~\ref{fig:shared_vs_original_length}(b) scans $10\leq T\leq10^5$. To separate narrow coherent peaks from a stable high-fidelity regime, we use the persistent threshold time $T_{0.99}^{\mathrm{st}}$, defined as the first sampled time after which every later point satisfies $P_{\mathrm{tar}}\geq0.99$. As summarized in Table~\ref{tab:shared_length_comparison}, the 39-site shared chain reaches this regime about $2.5$ times faster than the 38-site original chain, and the 59-site shared chain is about $18$ times faster than its 58-site counterpart. The 79-site shared chain remains above $0.99$ after approximately $3.73\times10^3/w$, whereas the original 78-site chain does not reach $0.99$ within $T\leq10^5/w$ and has only $P_{\mathrm{tar}}\simeq0.9803$ at the upper limit. The propagation used $24000$ midpoint time slices, and representative trajectories and threshold points were checked with $48000$ slices; the maximum discrepancy in the final population was $1.49\times10^{-5}$.

Because the common peak weak bond remains below the critical value, the original chain must rely on an edge-to-edge finite-size hybridization that decreases exponentially with length. The shared-boundary construction instead decomposes the same transport distance into two shorter segments coupled through a localized boundary mode and consequently retains larger effective couplings. The shared protocol therefore gives both higher fixed-time fidelity and shorter high-fidelity adiabatic times for all three resource-matched pairs, with an advantage that becomes larger over the three system sizes considered here.

\section{Discussions and conclusions}\label{sec:summary}

\subsection{Validity of the boundary-state reduction}

The central approximation in this work is the projection of an extended qubit chain onto a small set of localized boundary modes. For a single Rice--Mele chain, the relevant subspace is spanned by the left and right edge states; for two segments joined at one physical boundary qubit, it is spanned by the left, shared, and right boundary modes. This reduction is controlled when these modes remain identifiable and sufficiently separated from the bulk spectrum. If the followed branch strongly hybridizes with bulk states, the two- or three-state Hamiltonian no longer provides a quantitatively complete description, and the microscopic chain dynamics must be retained. The agreement with the full-chain simulations therefore serves as a check of the boundary-state approximation rather than as a consequence of the reduced model alone.

The same viewpoint exposes the principal tradeoff in the transfer protocols. Deeper localization suppresses overlap with the bulk, but it also reduces the finite-size coupling between distant boundary modes. In the dark-state protocol, the relevant gap is set by $\sqrt{g_1^2+g_2^2}$ and inherits the exponential length dependence of $g_1$ and $g_2$. The dark channel is consequently effective for the short segments studied above but becomes slow as the segments are lengthened. The on-site-potential-driven shared-boundary protocol is more favorable in the resource-matched comparison because the common node partitions the total distance into two shorter overlaps. Thus the shared boundary does not remove the localization--speed tradeoff; it changes the geometry of the effective channel so that larger couplings are retained over the same overall distance.

\subsection{Experimental considerations}

The proposal requires three operations in the single-excitation sector: preparation of an excitation at an outer boundary, time-dependent control of the exchange couplings and qubit frequencies, and site-resolved measurement at the end of the evolution. Gap-tunable flux qubits provide one implementation, while other superconducting-qubit architectures with tunable exchange and frequency control can realize the same Hamiltonian~\cite{Paauw2009,Schwarz2013,chen2014qubit,geller2015tunable}. Independent controls implement $v_1(t)$, $v_2(t)$ and, when needed, the boundary-sector potentials $\Delta_1(t)$, $\Delta_2(t)$, and $\Delta_3(t)$.

Experimental performance is set by the minimum instantaneous gap relative to the available coherence time and control bandwidth: rapid evolution produces nonadiabatic transitions, whereas long evolution accumulates relaxation and dephasing. In addition, calibrated boundary-mode energies are needed to maintain spectral isolation and, for the dark-state protocol, the three-mode resonance condition.

The final boundary occupation can be obtained through standard dispersive readout, in which a state-dependent shift of a detuned resonator is inferred from the phase of a reflected or transmitted probe~\cite{deac2008bias,beri2010topologically,romero2009microwave,mallet2009single}. A cavity-mediated circuit-QED array provides another route to the same effective exchange model when the cavity modes are eliminated in the dispersive regime~\cite{Devoret2013,Gu2017}.

\subsection{Conclusions}

We have formulated topological edge-state transfer as adiabatic evolution within an effective boundary-mode channel. In a single superconducting-qubit Rice--Mele chain, projection onto the two end modes produces an LZ Hamiltonian. Its detuning and finite-size hybridization explain the edge-to-edge passage and identify the spectral separation from bulk states as the relevant constraint for optimizing the protocol.

We then extended this construction to two Rice--Mele segments that share one physical boundary qubit. The three localized modes yield a general three-state Hamiltonian with microscopically determined energies $\Delta_i$ and couplings $g_i$. This channel supports a zero-energy dark-state passage and a continuous on-site-potential-driven passage. Full-chain simulations reproduce the reduced dynamics in the regime where the boundary modes remain isolated. The length comparison further shows that the shared-boundary geometry can reach high fidelity in a shorter time than a single uninterrupted chain under the same control bounds because it replaces one long, exponentially weak overlap by two shorter ones.

Boundary-mode projection relates the transfer dynamics directly to finite-size hybridization and the corresponding adiabatic gaps. In the shared-boundary geometry, replacing one long overlap by two shorter overlaps increases these gaps under the control bounds considered here and reduces the time required to reach stable high-fidelity transfer.

\begin{acknowledgments}
Y.X.L. acknowledges the support of the National Basic Research Program of
China Grant No. 2014CB921401 and the National Natural Science Foundation of
China under Grant No. 91321208. S.C. is supported by the National Key Research
and Development Program of China (2016YFA0300600), NSFC under Grants No. 11425419, No. 11374354 and No. 11174360.
\end{acknowledgments}

\appendix \label{appendix}

\section{Qubit chain with tunable couplings}\label{sec:chain}

The protocols considered in the main text require control of the staggered qubit frequencies and of the two alternating exchange couplings. We describe two standard routes to the required Hamiltonian. The first uses longitudinal frequency modulation to renormalize or parametrically activate exchange interactions. The second uses flux-tunable couplers between neighboring qubits. Both routes preserve excitation number and therefore lead to the same single-excitation Rice--Mele description used in the main text.

\subsection{Tunable coupling realized by longitudinal frequency modulation}

We enumerate the physical qubits as $j=1,\ldots,2L$, with $j=2n-1$ corresponding to $A_n$ and $j=2n$ to $B_n$. Let $v_0$ and $w_0$ denote the unmodulated intracell and intercell exchange amplitudes, respectively. A longitudinally modulated chain is described by
\begin{align}
	H_M(t)={}&
	\frac{1}{2}\sum_{j=1}^{2L}
	\left[\omega_j+f_j(t)\right]\sigma_j^z
	\nonumber\\
	&+v_0\sum_{\substack{j=1\\j\ {\rm odd}}}^{2L-1}
	\left(\sigma_j^+\sigma_{j+1}^-+\mathrm{H.c.}\right)
	\nonumber\\
	&+w_0\sum_{\substack{j=1\\j\ {\rm even}}}^{2L-1}
	\left(\sigma_j^+\sigma_{j+1}^-+\mathrm{H.c.}\right),
	\label{eq:5-1}
\end{align}
where
\begin{equation}
	f_j(t)=\epsilon_j\cos(\nu_j t+\phi_j).
	\label{eq:appendix_frequency_modulation}
\end{equation}
Here $\epsilon_j$, $\nu_j$, and $\phi_j$ are the modulation amplitude, frequency, and phase. Such longitudinal modulation can be implemented by flux control of the qubit transition frequency~\cite{goetz2018parity,wang201816}.

For clarity, we first set $\phi_j=0$ and define the dimensionless modulation amplitude $\eta_j=\epsilon_j/\nu_j$. The local rotating transformation is
\begin{equation}
	U_j(t)=
	\exp\left\{-\frac{i}{2}
	\left[\omega_jt+\eta_j\sin(\nu_jt)\right]\sigma_j^z\right\}.
	\label{eq:appendix_local_rotation}
\end{equation}
Under $U(t)=\prod_jU_j(t)$, the raising and lowering operators become
\begin{align}
	U_j^\dagger(t)\sigma_j^\pm U_j(t)
	={}&\sigma_j^\pm e^{\pm i\omega_jt}
	\exp\left[\pm i\eta_j\sin(\nu_jt)\right]
	\nonumber\\
	={}&\sigma_j^\pm e^{\pm i\omega_jt}
	\sum_{m=-\infty}^{\infty}
	\mathcal{J}_m(\pm\eta_j)e^{im\nu_jt},
	\label{eq:appendix_ladder_transform}
\end{align}
where $\mathcal{J}_m$ is a Bessel function of the first kind. This expression shows explicitly how the longitudinal drive generates exchange sidebands.

We first consider equal bare qubit frequencies, $\omega_j=\omega_0$, and a common modulation frequency, $\nu_j=\nu$. Retaining the static component in the rotating-wave approximation gives
\begin{align}
	H_1={}&
	\sum_{\substack{j=1\\j\ {\rm odd}}}^{2L-1}
	v_j^{\mathrm{eff}}
	\left(\sigma_j^+\sigma_{j+1}^-+\mathrm{H.c.}\right)
	\nonumber\\
	&+\sum_{\substack{j=1\\j\ {\rm even}}}^{2L-1}
	w_j^{\mathrm{eff}}
	\left(\sigma_j^+\sigma_{j+1}^-+\mathrm{H.c.}\right).
	\label{eq:5-2}
\end{align}
The effective alternating couplings are
\begin{align}
	v_j^{\mathrm{eff}}
	&=v_0\sum_{m=-\infty}^{\infty}
	\mathcal{J}_m(\eta_j)\mathcal{J}_m(\eta_{j+1})
	\nonumber\\
	&=v_0\mathcal{J}_0(\eta_j-\eta_{j+1}),
	\qquad j\ {\rm odd},
	\label{eq:appendix_same_frequency_v}\\
	w_j^{\mathrm{eff}}
	&=w_0\sum_{m=-\infty}^{\infty}
	\mathcal{J}_m(\eta_j)\mathcal{J}_m(\eta_{j+1})
	\nonumber\\
	&=w_0\mathcal{J}_0(\eta_j-\eta_{j+1}),
	\qquad j\ {\rm even}.
	\label{eq:appendix_same_frequency_w}
\end{align}
Thus the exchange magnitude on each bond is controlled by the difference between the modulation amplitudes applied to its two endpoint qubits. Site-selective or multitone modulation allows the odd and even bonds to be programmed independently. For a spatially uniform Rice--Mele segment, these values are chosen as $v_j^{\mathrm{eff}}=v(t)$ on intracell bonds and $w_j^{\mathrm{eff}}=w(t)$ on intercell bonds.

Frequency differences between neighboring qubits provide a second implementation based on resonant sideband matching~\cite{liu2006controllable}. Suppose the modulation frequencies are chosen so that one first-order sideband compensates the detuning on each addressed bond. After discarding all rapidly oscillating terms, the effective Hamiltonian retains the same alternating structure,
\begin{align}
	H_2={}&
	\sum_{\substack{j=1\\j\ {\rm odd}}}^{2L-1}
	v_j'
	\left(\sigma_j^+\sigma_{j+1}^-+\mathrm{H.c.}\right)
	\nonumber\\
	&+\sum_{\substack{j=1\\j\ {\rm even}}}^{2L-1}
	w_j'
	\left(\sigma_j^+\sigma_{j+1}^-+\mathrm{H.c.}\right).
	\label{eq:5-3}
\end{align}
For the sideband assignment used in the original construction, the retained amplitudes have the form
\begin{align}
	v_j'
	&=v_0\mathcal{J}_0(\eta_j)
	\mathcal{J}_1(\eta_{j+1})e^{i\chi_{v,j}},
	\qquad j\ {\rm odd},
	\label{eq:appendix_sideband_v}\\
	w_j'
	&=w_0\mathcal{J}_1(\eta_j)
	\mathcal{J}_0(\eta_{j+1})e^{i\chi_{w,j}},
	\qquad j\ {\rm even},
	\label{eq:appendix_sideband_w}
\end{align}
where $\chi_{v,j}$ and $\chi_{w,j}$ are set by the modulation phases. In an open one-dimensional chain these static bond phases can be removed by local phase redefinitions, so the effective couplings may be taken real. Parametrically activated exchange of this type has been demonstrated in coupled superconducting qubits and short qubit chains~\cite{wang201816,li2018perfect}. The staggered potential is implemented independently by choosing the slowly varying transition frequencies as
$\omega_{A,n}(t)=\omega_0+\Delta(t)$ and
$\omega_{B,n}(t)=\omega_0-\Delta(t)$.

\subsection{Tunable coupling via couplers}

A second route is to insert a flux-controlled coupler between each pair of neighboring qubits~\cite{liu2006controllable,Liu2007,Grajcar2006,Rigetti2005,Bertet2006,Niskanen2006,Hime2006,Niskanen2007,Harris2007,VanDerPloeg2007,Allman2010,Bialczak2011}. The coupler may operate as a classical nonlinear element, in which case its dynamical variable is adiabatically eliminated~\cite{Grajcar2006}, or as a quantum mode that is eliminated in a dispersive regime~\cite{liu2005measuring}. In both cases, the applied coupler flux controls the exchange interaction between its two neighboring qubits.

After eliminating the coupler degrees of freedom, the qubit-chain Hamiltonian can be written as
\begin{align}
	H_C(t)={}&
	\frac{1}{2}\sum_{n=1}^{L}
	\left[\widetilde{\omega}_{A,n}(t)\sigma_{A,n}^z
	+\widetilde{\omega}_{B,n}(t)\sigma_{B,n}^z\right]
	\nonumber\\
	&+\sum_{n=1}^{L}v_n(t)
	\left(\sigma_{A,n}^+\sigma_{B,n}^-+\mathrm{H.c.}\right)
	\nonumber\\
	&+\sum_{n=1}^{L-1}w_n(t)
	\left(\sigma_{B,n}^+\sigma_{A,n+1}^-+\mathrm{H.c.}\right).
	\label{eq:RM}
\end{align}
The renormalized frequencies $\widetilde{\omega}_{A,n}$ and $\widetilde{\omega}_{B,n}$ include the shifts generated when the couplers are eliminated. These shifts can be incorporated into the calibrated frequency controls. For a uniform Rice--Mele segment we choose
$v_n(t)=v(t)$, $w_n(t)=w(t)$,
$\widetilde{\omega}_{A,n}(t)=\omega_0+\Delta(t)$, and
$\widetilde{\omega}_{B,n}(t)=\omega_0-\Delta(t)$.
After removing the common frequency $\omega_0$ and restricting to the single-excitation sector, Eq.~(\ref{eq:RM}) reduces to Eq.~(\ref{Hamiltonian:Pumping}).

The same control architecture extends directly to the shared-boundary chain. Independent coupler biases generate $v_1(t),w_1(t),v_2(t),w_2(t)$, while calibrated qubit-frequency biases produce the boundary-sector potentials $\Delta_1(t),\Delta_2(t),\Delta_3(t)$. Because both implementations contain only longitudinal frequency terms and excitation-exchange interactions, the excitation-number conservation assumed throughout the main text is preserved.

\bibliographystyle{apsrev4-2}
\bibliography{library}
\end{document}